\documentclass[conference]{IEEEtran}
\IEEEoverridecommandlockouts

\usepackage{cite}
\usepackage{amsmath,amssymb,amsfonts}
\usepackage{algorithmic}
\usepackage{graphicx}
\usepackage{textcomp}
\usepackage{xcolor}

\usepackage{siunitx}
\usepackage{url}

\usepackage{booktabs}
\usepackage{multirow}
\usepackage{microtype}
\usepackage{stfloats}
\usepackage{comment}

\def\BibTeX{{\rm B\kern-.05em{\sc i\kern-.025em b}\kern-.08em
    T\kern-.1667em\lower.7ex\hbox{E}\kern-.125emX}}

\begin{document}

\title{HPRO: Hierarchical Progressive Reward Optimization via Preference Extraction for Emotional Text-to-Speech}

\author{
\IEEEauthorblockN{
Sihang Nie\textsuperscript{1,$\dagger$}, 
Xiaofen Xing\textsuperscript{1,$*$}, 
Rui Xing\textsuperscript{1,$\dagger$}, 
Haoming Li\textsuperscript{2}, 
Ruitong Xiao\textsuperscript{2}, \\
Jingyuan Xing\textsuperscript{1}, 
Baiji Liu\textsuperscript{1,3}, 
and Xiangmin Xu\textsuperscript{1,4}
}
\vspace{1.5mm}
\IEEEauthorblockA{\textsuperscript{1}South China University of Technology, \textsuperscript{2}Huya Inc., \\
\textsuperscript{3}Tongyi Fun Team, Alibaba Group,
\textsuperscript{4}Foshan University \\
xfxing@scut.edu.cn, bcshnie@mail.scut.edu.cn
}
}


\maketitle

\renewcommand{\thefootnote}{\fnsymbol{footnote}}
\footnotetext[2]{Work conducted when the author was intern at Huya Inc.}
\footnotetext[1]{Corresponding author.}
\renewcommand{\thefootnote}{\arabic{footnote}}

\begin{abstract}
Recently, Large Language Model (LLM)-based Text-to-Speech (TTS) models have achieved remarkable naturalness. However, the standard Supervised Fine-Tuning paradigm often converges to statistically averaged prosody, limiting emotional expressiveness. While preference-driven optimization offers a promising alternative, existing approaches suffer from two structural mismatches: information conflict, where content and emotion in a shared latent space produce conflicting gradients, leading to reward hacking and semantic degradation; and scale gap, where sparse sentence-level rewards struggle to guide dense frame-level generation. To overcome these challenges, we propose HPRO, a hierarchical progressive reward optimization framework. Within HPRO, we introduce the HD-Emo codec as a novel differentiable reward model to mitigate the information conflict. It extracts speech into distinct content and style preference tokens, structurally isolating emotional optimization from semantic content. Building upon this structured preference space, HPRO bridges the scale gap by progressively aligning frame-, word- and sentence-level objectives. Experiments demonstrate that HPRO significantly enhances emotional expressiveness, while effectively preserving linguistic intelligibility. The code and audio samples are publicly available at~\url{https://xxh333.github.io/hpro-demo/}. 
\end{abstract}


\section{Introduction}
In recent years, Large Language Model (LLM)-based approaches have driven remarkable advancements in speech synthesis. By treating Text-to-Speech (TTS) as a next-token prediction task within an LLM framework~\cite{seedtts,fireredtts2,qwen3tts}, models have achieved unprecedented naturalness and intelligibility. 
Based on this foundation, emotional TTS~\cite{indextts2,emovoice,wescon} has also made notable progress, enabling models to achieve basic emotional control and generate speech with specific affective states. Despite these advances, synthesizing highly expressive and authentic emotional speech remains a challenge. 
The mainstream Supervised Fine-Tuning (SFT) paradigm is inherently constrained by a ``regression to the mean'' effect. Specifically, the cross-entropy (CE) objective forces the models to approximate the conditional average of the training data, resulting in flattened prosody that lacks the nuance and vigor of authentic human expression~\cite{prosodytts}.

\begin{figure}[t]
  \centering
  \includegraphics[width=\linewidth]{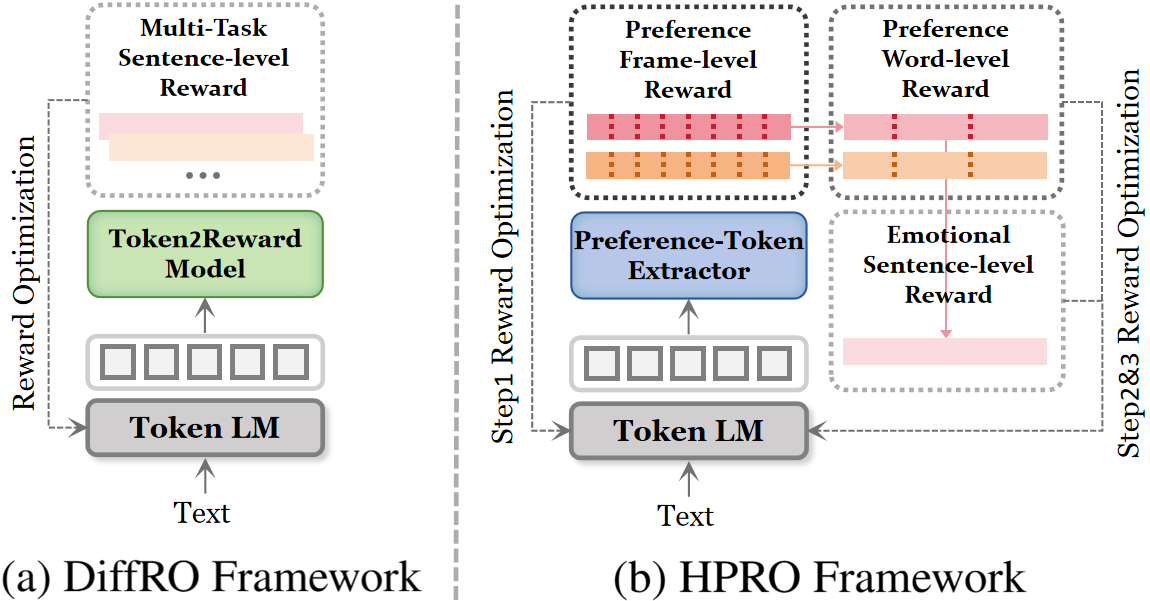}
  \caption{Motivation. (a) DiffRO Framework. Single-scale reward optimization for monolithic speech representation. (b) HPRO Framework. Hierarchical progressive reward optimization for structured preference spaces.}
  \label{fig:mot}
\end{figure}

To overcome the limitations of SFT and better align synthesis with human affective perception, preference-driven optimization has been increasingly explored in emotional TTS. 
Early reinforcement learning (RL) efforts, such as i-ETTS~\cite{ietts}, utilized speech emotion recognition (SER) rewards to optimize generation policies via policy gradients. 
Subsequently, a line of work~\cite{emodpo,emolipo} adopted direct preference optimization (DPO)-style objectives, which are learned from constructed paired preference data without explicit reward modeling. 
More recent studies~\cite{emorltts,rlaif,glmtts} have explored group relative policy optimization (GRPO) to design emotion-aware reward functions and perform policy optimization toward higher emotional intensity. 
In parallel, DiffRO~\cite{diffro} proposed a differentiable reward framework that predicts reward values directly from speech tokens rather than synthesized waveforms, allowing direct backpropagation-based optimization of LLM's parameters without relying on conventional RL rollouts, as shown in Fig.~\ref{fig:mot}(a).

However, directly transplanting these paradigms to emotional TTS poses significant challenges. In practice, such optimization is prone to reward hacking~\cite{rrpo}, where models maximize emotional scores at the catastrophic expense of intelligibility. We attribute this to two structural mismatches: 
1) \textbf{Information Conflict}: In monolithic speech spaces, semantic content and emotional style share the same latent representation. Consequently, global reward maximization becomes a conflicting optimization objective~\cite{noreward}, where intensifying emotion often disrupts acoustic structures that contain linguistic content. 
2) \textbf{Scale Gap}: There is a severe granularity mismatch between sparse emotional rewards at the sentence-level and the dense frame-level nature of speech generation~\cite{w3ar}. This disconnect leads to a credit assignment dilemma, as the model lacks explicit mechanisms to focus on emotionally significant segments within dense generation.
In particular, RRPO~\cite{rrpo} mitigates reward hacking by improving the robustness of the reward model through hybrid regularization. However, this strategy primarily improves reward reliability without fundamentally addressing the two mismatches.

\begin{figure*}[htbp]
    \centering
    \includegraphics[width=14.5cm]{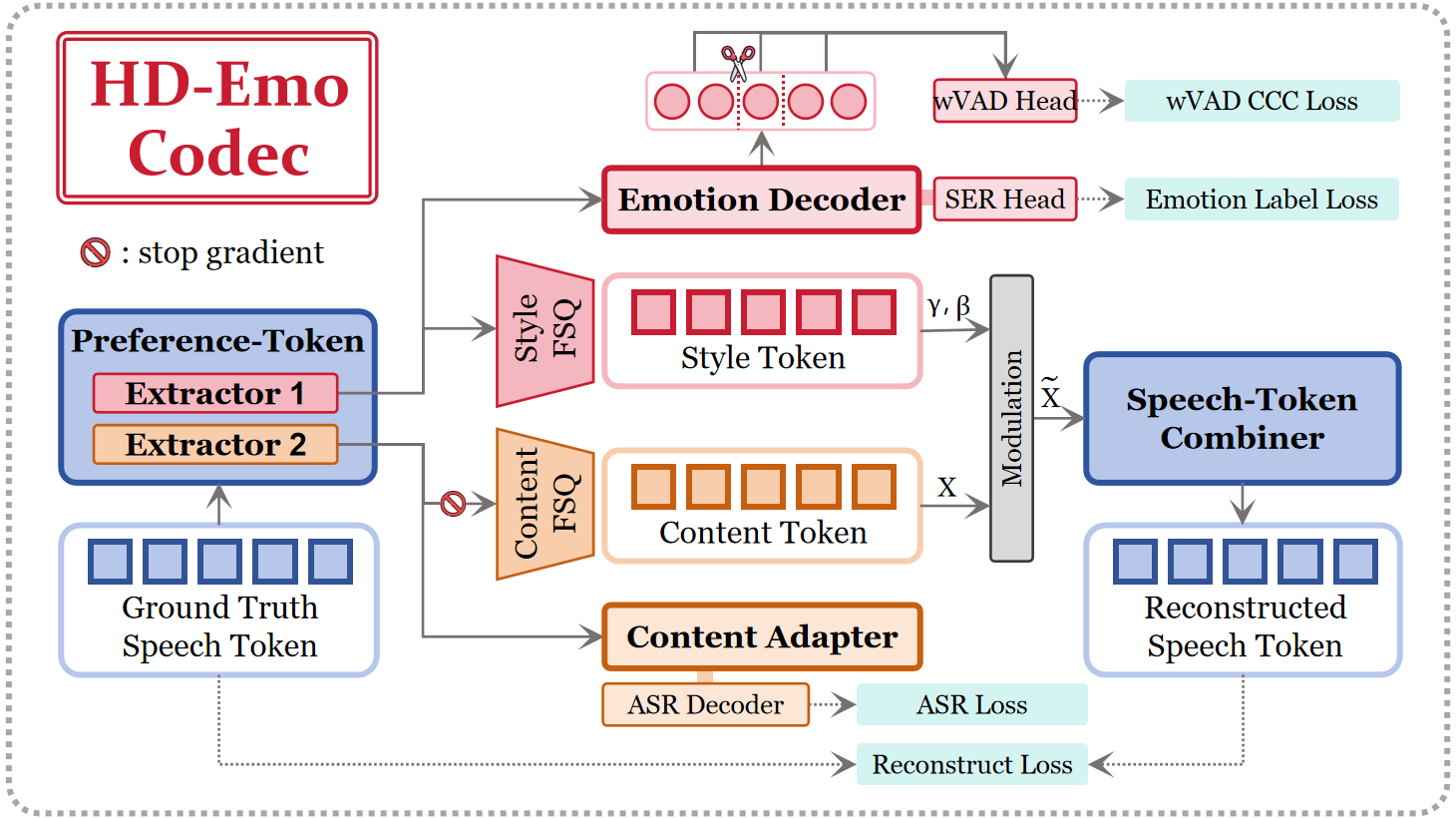}
    \caption{Overview of the HD-Emo Codec. Monotonic speech tokens are processed by dual preference extractors with FSQ bottlenecks to obtain content and style preference tokens. The two streams are respectively supervised by ASR and hierarchical emotional objectives (SER and wVAD), and subsequently fused via dynamic feature modulation for speech token reconstruction.}
  \label{fig:codec}
\end{figure*}

To resolve these issues, we propose \textbf{HPRO}, a \textbf{H}ierarchical \textbf{P}rogressive \textbf{R}eward \textbf{O}ptimization framework, as shown in Fig.~\ref{fig:mot}(b). 
Within this framework, we first address the information conflict by introducing the HD-Emo codec as a novel differentiable reward model. Revisiting the token codec design in HD-PPT~\cite{hdppt}, we adapt its architecture to serve as a structured reward interface. This preference-oriented token codec projects speech tokens into distinct content and style preference subspaces, ensuring prosodic integrity via reconstruction while structurally isolating stylistic optimization from semantic content.
Specifically, we employ automatic speech recognition (ASR) supervision on content tokens to maintain linguistic accuracy, and speech emotion recognition (SER) objectives on style tokens for global affective alignment. Furthermore, inspired by EmoSphere-TTS~\cite{emosphere}, we incorporate word-level Valence–Arousal–Dominance (wVAD) constraints to provide fine-grained emotional guidance. 
Building upon this structured preference space, HPRO bridges the scale gap through a progressive optimization mechanism. Instead of relying solely on sparse feedback, this approach constructs a continuous gradient bridge—spanning dense frame-level supervision, word-level boundary constraints, and global sentence-level objectives—thereby ensuring robust optimization.

Our main contributions are summarized as follows:
\begin{itemize}
\item We propose the HPRO framework, which employs a progressive gradient path to effectively bridge the scale gap between sparse rewards and dense generation.
\item Within HPRO, we introduce the HD-Emo codec as a differentiable reward model, which extracts distinct preference tokens to structurally mitigate the information conflicts. 
\item Experiments demonstrate that HPRO enhances fine-grained emotional expressiveness while effectively preventing semantic degradation.
\end{itemize}

\section{Methodology}
\subsection{Differentiable Reward Modeling via HD-Emo Codec}
\label{ssec:codec}

Serving as the differentiable reward model for our HPRO framework, the HD-Emo codec projects speech tokens into specific preference subspaces, as shown in Fig.~\ref{fig:codec}. 
The codec processes discrete speech tokens (extracted by the CosyVoice2 tokenizer~\cite{cosyvoice2}) through two preference token extractors with identical architectures but unshared parameters. 
Both streams use finite scalar quantization (FSQ)~\cite{fsq}, with a straight-through estimator (STE) during codec pretraining. FSQ imposes a strict information bottleneck, compressing latent representations into discrete content-preference tokens $T_c$ and style-preference tokens $T_s$.

To strictly anchor $T_c$ to linguistic semantics, latent representations prior to quantization $Z_c$ are fed into a content adapter for transcription prediction. To ensure robust linguistic extraction, we employ an ASR objective consistent with Whisper~\cite{whisper}, using an ASR decoder initialized with the pre-trained Whisper-medium decoder\footnote{\url{https://huggingface.co/openai/whisper-medium}}. Given the ground-truth text transcription sequence $Y = (y_1, y_2, \dots, y_M)$, the ASR loss is formulated as the auto-regressive negative log-likelihood of the target text tokens:
\begin{equation}
\mathcal{L}_{ASR} = - \sum_{j=1}^{M} \log P(y_j \mid y_{<j}, Z_c)
\end{equation}
where $y_{<j}$ denotes the preceding text tokens. Crucially, to prevent acoustic leakage—where the extractor captures prosodic details to aid reconstruction—we apply a stop-gradient mechanism. By detaching the gradient flow from the reconstruction path, the content extractor is updated exclusively by the ASR supervision, ensuring strict semantic alignment.

\begin{figure*}[htbp]
    \centering
    \includegraphics[width=\linewidth]{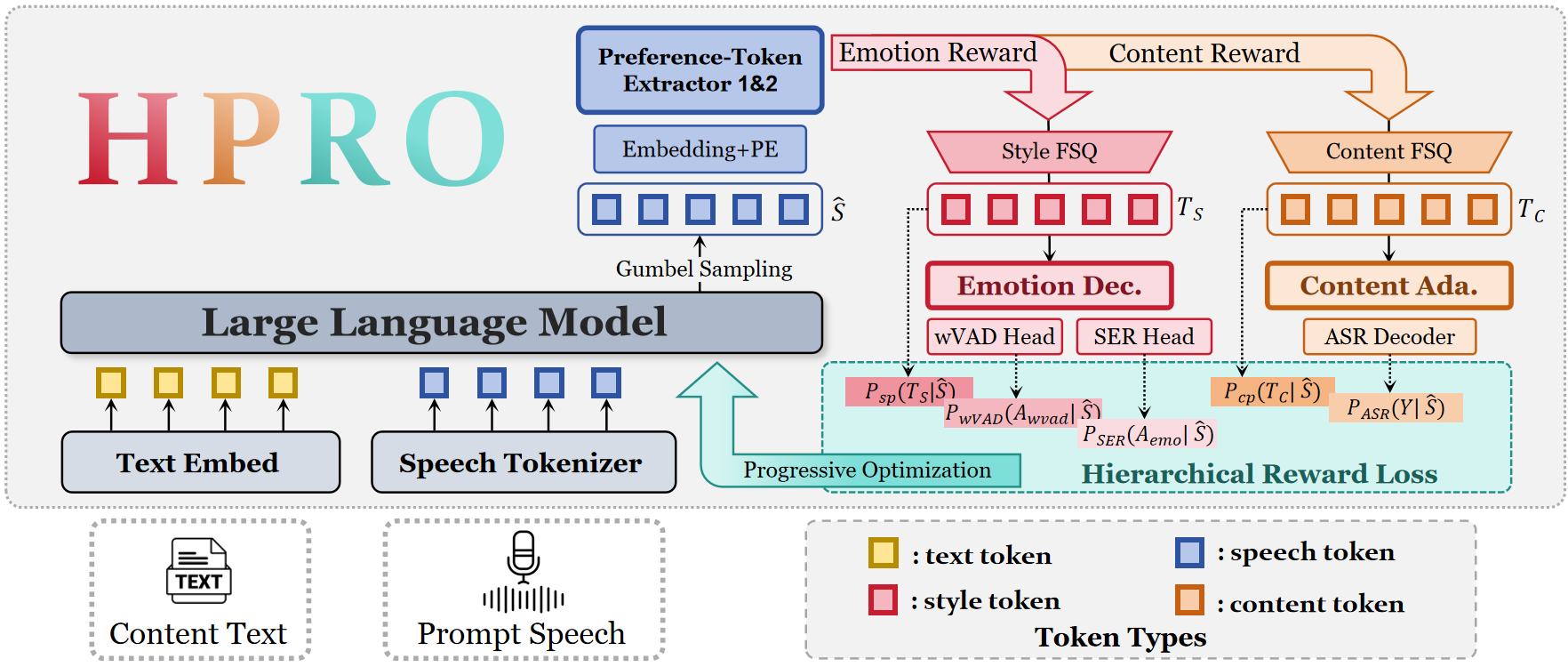}
    \caption{Overview of the HPRO framework. The LLM generates differentiable speech tokens via Gumbel-Softmax, which are mapped by HD-Emo codec into preference spaces. Hierarchical frame-, word- and sentence-level rewards are progressively applied to update the LLM.}
    \label{fig:llm}
\end{figure*}

In contrast, $T_s$ is optimized to capture expressive features via hierarchical supervision. At the sentence level, a pre-trained emotion2vec~\cite{emotion2vec} model\footnote{\url{https://huggingface.co/emotion2vec/emotion2vec_plus_large}} provides a soft emotion distribution $p$, which supervises the predicted distribution $\hat{p}$ from the emotion decoder through the CE loss over $C$ emotion categories:
\begin{equation}
\mathcal{L}_{SER} = - \sum_{i=1}^{C} p_i \log \hat{p}_i
\end{equation}

For fine-grained stylistic control, we employ the Montreal Forced Aligner (MFA)~\cite{mfa} tool~\footnote{\url{https://montreal-forced-aligner.readthedocs.io/en/latest/index.html}} to temporally align text with audio. Based on the aligned boundaries, we derive wVAD trajectories using a pre-trained Wav2vec2-ft model~\footnote{\url{https://huggingface.co/audeering/wav2vec2-large-robust-12-ft-emotion-msp-dim}}. 
To account for transitional prosody, we compute wVAD metrics over a contextual window including one word on each side of the target word. We employ the Concordance Correlation Coefficient (CCC)~\cite{ccc} to measure the agreement between the target $v$ and the prediction $\hat{v}$:
\begin{equation}
\text{CCC}(v, \hat{v}) = \frac{2 \sigma_{v \hat{v}}}{\sigma_v^2 + \sigma_{\hat{v}}^2 + (\mu_v - \mu_{\hat{v}})^2}
\end{equation}
where $\mu$ and $\sigma^2$ denote the mean and variance, respectively, and $\sigma_{v \hat{v}}$ represents the covariance. Consequently, the word-level loss is defined to enforce consistency across the three dimensions:
\begin{equation}
    \mathcal{L}_{word} = \sum_{k \in \{V,A,D\}} (1 - \text{CCC}(v_k, \hat{v}_k))
\end{equation}

Finally, to effectively integrate the extracted preference tokens for reconstruction, we apply a dynamic feature modulation mechanism inspired by Emo-FiLM~\cite{emofilm} before passing them to a speech token combiner. Specifically, the content representation $X$ (derived from $T_c$) is modulated by scaling $\gamma$ and shifting $\beta$ factors projected from $T_s$, formulated as follows:

\begin{equation}
    \widetilde{X} = X \odot \gamma + \beta
\end{equation}

Reconstruction is optimized through the CE loss. This design preserves the integrity of prosodic information within the frame-level preference tokens while maintaining structural isolation of content and style.

\subsection{HPRO}
To address structural mismatches in differentiable optimization for emotional TTS, we propose the HPRO framework, as shown in Fig.~\ref{fig:llm}. We utilize the pre-trained HD-Emo codec to extract speech tokens into distinct preference spaces, effectively mitigating the information conflict. Building upon this structured space, we bridge the scale gap by organizing supervision at the frame-, word- and sentence-levels. To balance these hierarchical rewards, we adopt a progressive strategy that gradually introduces higher-level objectives during training.

\subsubsection{Hierarchical Reward Formulation}
\label{ssec:reward_design}
The hidden states of the LLM are projected onto the discrete speech token space and sampled via the straight-through Gumbel-Softmax operation to obtain a differentiable token sequence $\hat{S}$. 
The frozen HD-Emo codec then maps $\hat{S}$ into structured preference spaces, producing the generated pre-FSQ features $(\hat{Z}_c,\hat{Z}_s)$ and preference tokens $(\hat{T}_c,\hat{T}_s)$ alongside their supervisory predictions. This mechanism allows gradients to propagate back to the LLM through the relaxed token representations.

Based on these preference representations, we define hierarchical reward functions at three levels: 

\textbf{Frame-level reward.} To provide dense, frame-level acoustic supervision, we align the generated pre-quantization representations ($\hat{Z}_c, \hat{Z}_s$) directly with the ground-truth discrete preference tokens ($T_c, T_s$). This alignment is performed in the FSQ latent dimension using $L_1$ regression losses:
\begin{equation}
\mathcal{L}_{cp} = \|\hat{Z}_c - T_c\|_1, \quad \mathcal{L}_{sp} = \|\hat{Z}_s - T_s\|_1
\end{equation}

\textbf{Word-level reward.} Using MFA boundaries, we impose a wVAD CCC loss $\mathcal{L}_{wVAD}$ to match predicted wVAD trajectories with target signals. To preserve semantic consistency, we incorporate a CE ASR loss $\mathcal{L}_{ASR}$ to penalize lexical deviations. The specific formulations of these losses strictly follow the HD-Emo codec design described in Section~\ref{ssec:codec}.

\textbf{Sentence-level reward.} The predicted emotion distribution is aligned with target soft labels using a CE loss $\mathcal{L}_{SER}$ to ensure global affective consistency.

We further retain a token-wise categorical KL divergence loss $\mathcal{L}_{KL}$ on the original speech token distribution to regularize the LLM outputs. The overall objective is formulated as follows:
\begin{equation}
\mathcal{L}_{total} = \sum_{i} \lambda_i \mathcal{L}_i,
\end{equation}
where $i \in \{KL, cp, sp, wVAD, ASR, SER\}$, and $\lambda_i$ controls the relative contribution of each supervision term.

\subsubsection{Progressive Optimization Strategy}
\label{ssec:progressive}
To ensure stable convergence during hierarchical reward learning, we introduce a progressive optimization strategy. This approach gradually expands the scope of supervision from local preference alignment to global affective objectives, effectively preventing early-stage reward hacking.

\textbf{Stage I: Frame-level warm-up.}
We initially optimize only the frame-level preference alignment losses  ($\mathcal{L}_{cp}$ and $\mathcal{L}_{sp}$) alongside the KL regularization term. This stage grounds the LLM outputs within the structured preference space before introducing higher-level semantic or emotional objectives. The loss weights are set to $\lambda_{KL}=0.05$, $\lambda_{sp}=2$, and $\lambda_{cp}=1$. The Gumbel temperature is initialized to $\tau=2$ to provide smooth gradients during the initial training phase.

\textbf{Stage II: Word-level refinement.}
We then introduce word-level supervision through $\mathcal{L}_{wVAD}$ and $\mathcal{L}_{ASR}$ to refine local emotional trajectories while strictly preserving semantic consistency. The loss weights are adjusted to $\lambda_{KL}=0.02$, $\lambda_{sp}=2$, $\lambda_{cp}=1$, $\lambda_{ASR}=5$, and $\lambda_{wVAD}=1$. The Gumbel temperature is annealed to $\tau=1$, allowing for sharper token selection as supervision becomes more structured.

\textbf{Stage III: Sentence-level alignment.}
Finally, we incorporate the sentence-level emotion classification loss $\mathcal{L}_{SER}$ to unify the global affective style. Building upon Stage II, we set $\lambda_{SER}=0.5$ and further anneal the temperature to $\tau=0.8$. This encourages confident discrete token generation under comprehensive hierarchical supervision.

This progressive scheme establishes a stable optimization path from dense token-level alignment to global affective consistency, directly bridging the scale gap to effectively mitigate reward domination and semantic degradation.

\section{Experiments}
\subsection{Experimental Setup}
\subsubsection{Datasets and Baselines}
We evaluate our approach on three datasets. LibriSpeech~\cite{librispeech} (960h) is utilized to provide foundational ASR supervision for semantic extraction. LSSED~\cite{lssed} (206h), a large-scale categorical SER dataset, and EmoVoice-DB~\cite{emovoice} (40h), a highly expressive emotional TTS corpus, serve as the primary resources for emotional modeling, TTS training, and evaluation. Both emotional datasets are further incorporated into the ASR supervision during codec training to ensure robust TTS alignment under expressive prosody. We adopt the test sets of LSSED and EmoVoice-DB for evaluation across all experiments. 
We compare our method with strong zero-shot TTS baselines: \textbf{CosyVoice2}, \textbf{CosyVoice3}, \mbox{\textbf{IndexTTS2}}, and \textbf{HD-PPT}. For a fair comparison, HD-PPT is adapted from its original instructional framework to the zero-shot TTS setting. HPRO is implemented on top of the CosyVoice2 backbone. Furthermore, since the official implementation of DiffRO~\cite{diffro} is not publicly available, we simulate its single-scale reward optimization paradigm within our framework. The direct comparison against this simulated DiffRO baseline is explicitly detailed in the ablation study.

\subsubsection{Implementation and Training Details}
The HD-Emo codec consists of an 8-layer conformer~\cite{conformer} for dual preference-token extractors and an 8-layer autoregressive transformer as the speech-token combiner. The FSQ codebook sizes are 1296 for content tokens and 64 for style tokens. Training proceeds in two stages. The content branch is first pre-trained on LibriSpeech with ASR supervision and then continued on the emotional datasets. Subsequently, it is frozen while the remaining modules are optimized on LSSED and EmoVoice-DB. The codec is trained using the Adam~\cite{adam} optimizer for 100 epochs on 8 NVIDIA RTX 4090 GPUs with a learning rate of $1 \times 10^{-4}$.
The underlying LLM of our HPRO framework is built on the Qwen2.5-0.5B~\cite{qwen2} architecture. During the preference-driven optimization, the LLM is optimized using the Adam optimizer with a learning rate of $1 \times 10^{-5}$. The optimization process strictly follows the three-stage progressive strategy detailed in Section~\ref{ssec:progressive}, where the hierarchical reward constraints and the Gumbel temperature are dynamically adjusted to ensure stable convergence from dense token-level alignment to global affective consistency.

\subsubsection{Evaluation Metrics}
All evaluations are conducted under a zero-shot TTS setting, where each test utterance is synthesized using a randomly selected reference from the same speaker.
For subjective evaluation, we select a total of 90 emotionally balanced utterances and invite 18 participants to rate the samples on a 5-point Likert scale. 
The subjective metrics include \textbf{MOS-N} (naturalness) and \textbf{MOS-E} (consistency between emotion and semantic content). 
The objective metrics include \textbf{WER} (word error rate, computed via whisper-large-v3~\footnote{\url{https://huggingface.co/openai/whisper-large-v3}}), \textbf{wVAD-CCC}, and \textbf{EMO-SIM} (both consistent with our hierarchical emotional reward design), as well as \textbf{DNSMOS} (perceptual quality score predicted by the DNSMOS P.835 model~\footnote{\url{https://github.com/microsoft/DNS-Challenge/tree/master/DNSMOS}}). 
Notably, while HPRO optimizes the LLM within the discrete preference token space of the HD-Emo codec, these objective metrics are computed on the final synthesized waveforms using external models. This architectural mismatch prevents direct metric optimization and avoids evaluation circularity.
In all subsequent tables, subjective metrics are reported with their standard deviations. Furthermore, bold text indicates the best performance, and underlined text denotes the second-best. For transparency, detailed metric scores for individual audio samples are provided on our demo page.

\begin{table*}[!t]
\caption{Performance comparison on the LSSED and EmoVoice-DB test sets. TokenRecon reconstructs audio directly from ground-truth speech tokens as an acoustic reconstruction reference.}
\label{tab:main_results}
\centering
\resizebox{16.5cm}{!}{
\begin{tabular}{@{}lcccccc@{}}
\toprule
\multirow{2}{*}{\textbf{Model}} & \multicolumn{2}{c}{\textbf{Subjective}} & \multicolumn{4}{c}{\textbf{Objective}} \\
\cmidrule(lr){2-3} \cmidrule(lr){4-7}
& \textbf{MOS-N$^*$~↑} & \textbf{MOS-E$^*$~↑} & \textbf{\hspace{0.5em}WER~↓\hspace{0.5em}} & \textbf{wVAD-CCC~↑} & \textbf{EMO-SIM~↑} & \textbf{\hspace{0.3em}DNSMOS~↑\hspace{0.5em}} \\ 
\midrule
TokenRecon & - & - & 7.34\% & 0.570 & 0.775 & 3.58 \\
\midrule
CosyVoice2~\cite{cosyvoice2} & 4.094 $\pm$ 0.257 & 3.530 $\pm$ 0.402 & 5.45\% & 0.307 & 0.613 & \textbf{3.76} \\
CosyVoice3~\cite{cosyvoice3} & \underline{4.137 $\pm$ 0.285} & 3.538 $\pm$ 0.343 & \underline{4.90\%} & 0.275 & 0.611 & 3.72 \\
IndexTTS2~\cite{indextts2} & 4.026 $\pm$ 0.238 & \textbf{3.692 $\pm$ 0.213} & 6.74\% & 0.293 & 0.526 & 3.53 \\
HD-PPT~\cite{hdppt} & 4.068 $\pm$ 0.248 & 3.547 $\pm$ 0.328 & 4.92\% & \underline{0.323} & \underline{0.646} & \underline{3.75} \\
\midrule
\textbf{HPRO} & \textbf{4.171 $\pm$ 0.318} & \underline{3.650 $\pm$ 0.347} & \textbf{4.02\%} & \textbf{0.339} & \textbf{0.672} & 3.73 \\
\bottomrule
\multicolumn{7}{@{}l}{\rule{0pt}{3ex}\footnotesize $^*$ MOS $\pm$ denotes standard deviation.} \\
\end{tabular}}
\end{table*}

\subsection{Comparison with Baselines}
The comparative results on the LSSED and EmoVoice-DB test sets are presented in Table~\ref{tab:main_results}. TokenRecon reconstructs audio from ground-truth speech tokens and prompt speech. It serves as a reconstruction reference rather than a WER upper bound. As its subjective quality is inherently equivalent to the original human recordings, we omit its subjective evaluation and solely report its objective metrics to represent the theoretical ceiling.

In subjective evaluations, HPRO achieves the highest MOS-N and the second-highest MOS-E, demonstrating a superior balance between naturalness and emotional expressiveness. Although IndexTTS2 obtains the highest MOS-E score, this prominent emotional perceptibility stems from its reliance on text-predicted emotion probabilities mapped to predefined embeddings. This explicit mapping tends to produce overly intense yet stereotyped emotional expressions. Consequently, this coarse-grained approach lacks precise acoustic alignment with the target reference, a limitation clearly evidenced by its lowest MOS-N score and sub-optimal objective metrics.

In contrast, HPRO delivers a comprehensively superior and balanced performance across all objective metrics. Notably, it attains the lowest WER (4.02\%), proving that extracting distinct preference tokens via the HD-Emo codec successfully mitigates the information conflict, thereby preventing semantic degradation. Simultaneously, HPRO achieves the highest wVAD-CCC (0.339) and EMO-SIM (0.672). These results confirm that our hierarchical progressive optimization effectively bridges the scale gap, capturing intricate emotional nuances and fine-grained prosody without compromising linguistic intelligibility.

\begin{table}[t]
\caption{Ablation study on the progressive optimization strategy. Reward constraints are incrementally introduced on top of the CosyVoice2-SFT baseline.}
\label{tab:ablation_train}
\centering
\resizebox{\linewidth}{!}{%
\begin{tabular}{@{}lcccc@{}}
\toprule
\textbf{Model} & \textbf{WER~↓} & \textbf{wVAD-CCC~↑} & \textbf{EMO-SIM~↑} & \textbf{DNSMOS~↑} \\  \midrule
CosyVoice2-SFT & 5.42\% & 0.297 & 0.641 & 3.63 \\
\hspace{0.9em}+Frame & 4.85\% & 0.332 & 0.650 & \underline{3.71} \\
\hspace{1.8em}+Word & \textbf{3.99\%} & \textbf{0.350} & \underline{0.653} & 3.70 \\
\hspace{2.7em}+Sentence & \underline{4.02\%} & \underline{0.339} & \textbf{0.672} & \textbf{3.73} \\ \bottomrule
\end{tabular}
}
\end{table}

\subsection{Ablation on Hierarchical Preference Training}
Table~\ref{tab:ablation_train} validates the efficacy of our progressive optimization strategy.
Incorporating frame-level supervision (+Frame) consistently outperforms the SFT baseline, confirming that dense alignment on discrete preference tokens provides a robust acoustic foundation. This step is crucial, as it effectively initiates the gradient bridge from sparse rewards to dense frame-level generation, preventing the model from losing basic speech structures.

Building upon this acoustic foundation, introducing word-level constraints (+Word) yields the best WER and wVAD-CCC. This suggests that boundary-aware supervision successfully refines local prosodic variations, acting as an intermediate structural anchor that ensures precise temporal alignment between linguistic content and emotional expression.

Finally, adding the sentence-level objective (+Sentence) significantly boosts global emotion scores (EMO-SIM), although with a marginal decline in wVAD-CCC and WER. This minor divergence is expected and reveals an inherent tension in multi-scale generation: enforcing a unified global emotional category tends to slightly smooth out fine-grained, word-level prosodic fluctuations. It reflects the fundamental challenge of integrating global affective gradients with local acoustic constraints within the LLM's unified token generation space. Despite this minor local compromise, the full configuration achieves the highest global expressiveness and perceptual quality. Overall, the progressive integration of these hierarchical constraints yields substantial improvements over the baseline across all dimensions. This explicitly demonstrates that HPRO effectively navigates the optimization landscape, significantly mitigating the scale gap between dense local generation and sparse global emotional control.

\begin{table*}[!t]
\caption{Ablation study on individual reward components (trained in a non-progressive manner). Notably, the w/o frame\&wvad variant explicitly simulates the single-scale global reward paradigm of DiffRO.}
\label{tab:ablation_reward}
\centering
\resizebox{12cm}{!}{%
\begin{tabular}{@{}lcccc@{}}
\toprule
\textbf{Model} & \textbf{WER~↓} & \textbf{wVAD-CCC~↑} & \textbf{EMO-SIM~↑} & \textbf{DNSMOS~↑} \\  \midrule
w/o content & 13.61\% & 0.285 & 0.584 & 3.59 \\
w/o emotion & \textbf{3.80\%} & 0.295 & 0.637 & \textbf{3.78} \\
w/o frame & 4.97\% & \underline{0.333} & 0.608 & 3.68 \\
w/o wvad & 4.10\% & 0.310 & 0.659 & \underline{3.75} \\ 
w/o frame\&wvad (DiffRO) & 4.35\% & 0.315 & \underline{0.662} & 3.73 \\ \midrule
\textbf{HPRO} & \underline{4.02\%} & \textbf{0.339} & \textbf{0.672} & 3.73 \\ \bottomrule
\end{tabular}
}
\end{table*}

\subsection{Ablation on Reward Design}
Table~\ref{tab:ablation_reward} analyzes the contribution of distinct preference rewards. To strictly isolate the impact of individual reward components, all models in this specific ablation study are trained in a non-progressive manner (applying all specified rewards simultaneously from the start).

First, we examine the basic semantic and affective constraints. Removing all content-related objectives (\textit{w/o content}) leads to severe semantic degradation, with WER surging to 13.61\%. Conversely, removing emotion-related rewards (\textit{w/o emotion}) achieves the lowest WER (3.80\%) but severely lacks emotional expressiveness. This extreme trade-off perfectly illustrates the inherent information conflict: unconstrained emotional optimization inevitably disrupts acoustic structures containing linguistic content.

Next, we evaluate the hierarchical granularities. Omitting frame-level supervision (\textit{w/o frame}) causes a noticeable increase in WER (4.97\%) and a drop in EMO-SIM (0.608). This demonstrates that dense alignment provides an indispensable acoustic foundation. Similarly, removing word-level constraints (\textit{w/o wvad}) leads to declines in both wVAD-CCC (0.310) and EMO-SIM (0.659), proving its effectiveness in capturing fine-grained emotional fluctuations.

Most importantly, the configuration omitting both frame- and word-level rewards (\textit{w/o frame \& wvad}) essentially simulates the single-scale global reward paradigm of DiffRO. While it maintains decent global emotion (EMO-SIM of 0.662), it suffers from a degradation in WER (4.35\%) and wVAD-CCC (0.315). This contrast explicitly highlights the limitations of monolithic global rewards in balancing semantics and style. While the DiffRO-style baseline still struggles with this trade-off, the full HPRO configuration achieves the highest emotional expressiveness alongside a remarkably low WER (4.02\%). This concurrent optimization of emotion and semantics explicitly demonstrates that our framework effectively mitigates the information conflict.

\section{Conclusion}
In this work, we propose the HPRO framework to address the information conflict and scale gap inherent in preference-driven optimization for emotional TTS. To mitigate the semantic degradation caused by optimization within a monolithic latent space, we introduce the HD-Emo codec. It extracts speech tokens into distinct preference subspaces, structurally isolating affective optimization from semantic content. Furthermore, to alleviate the credit assignment challenges of single-scale global rewards, HPRO establishes a continuous gradient bridge that progressively integrates dense frame-level, word-level, and sentence-level objectives. Extensive experiments demonstrate that HPRO successfully mitigates both structural mismatches, achieving superior global and fine-grained emotional expressiveness while preserving high linguistic intelligibility, explicitly outperforming monolithic reward baselines.

In the future, we plan to explore the broader applicability of HPRO's hierarchical reward mechanism. Specifically, we aim to extend the word-level constraints from affective attributes to diverse stylistic features, transitioning them from explicit supervisory signals into learned intermediate representations. This will pave the way for investigating its potential in fine-grained word-level control for controllable TTS and advanced reward optimization for spoken dialogue models.

\clearpage
\bibliographystyle{IEEEtran}
\bibliography{mybib,paper}

\end{document}